\RequirePackage[l2tabu]{nag}

\documentclass[12pt,english]{article}

\usepackage[T1]{fontenc}
\usepackage[charter]{mathdesign} 

\usepackage{microtype}

\usepackage[longnamesfirst]{natbib}
\usepackage{url}
\makeatletter
\def\url@leostyle{%
    \def\UrlFont{\sf}}{\def\UrlFont{\small\ttfamily}}
\makeatother
\usepackage{geometry}
\usepackage[font=small,labelfont=bf]{caption}

\usepackage[hidelinks,breaklinks]{hyperref}

\usepackage{authblk}

\usepackage{enumitem}

\usepackage{xfrac}

\usepackage{graphicx}
\usepackage{babel}
\usepackage{amsmath}

\numberwithin{equation}{section}

\title{Reconsidering No-Go Theorems from a Practical
  Perspective\thanks{\textbf{Note:} this is the submitted version of
    an article that is to appear in \emph{The British Journal
    for the Philosophy of Science}, published by \emph{Oxford
    University Press.} Changes have been made to this article since it
    was submitted for publication. These improvements are mostly
    clarificatory in nature and do not substantially modify the
    structure or argument of the paper. They are primarily found in
    \textsection \ref{sec:gendisc} (which has been expanded), and at
    the end of \textsection \ref{sec:sims}. When citing this paper,
    please refer to the published version.
  }$\mbox{ }^,$\thanks{%
    Thanks to Sam Fletcher for his helpful comments and suggestions on
    an earlier draft of this paper. Thanks also to my audiences at
    the universities of Bristol (especially Ryan Samaroo), Florence,
    Helsinki, and Oxford for fruitful discussion. Thanks, finally, to
    the Alexander Humboldt Foundation, whose financial support has
    made this project possible.%
  }%
}

\author{Michael E. Cuffaro}
\affil{{\small Ludwig-Maximilians-Universit\"at M\"unchen,\\Munich
  Center for Mathematical Philosophy}}

\date{}

\begin{document}

\maketitle

\thispagestyle{empty}

\begin{abstract}
I argue that our judgements regarding the locally causal models which
are compatible with a given quantum no-go theorem implicitly depend,
in part, on the context of inquiry. It follows from this that certain
no-go theorems, which are particularly striking in the traditional
foundational context, have no force when the context switches to a
discussion of the physical systems we are capable of building with the
aim of classically reproducing quantum statistics. I close with a
general discussion of the possible implications of this for our
understanding of the limits of classical description, and for our
understanding of the fundamental aim of physical investigation.
\end{abstract}

\section{Introduction}
\label{sec:intro}

Bell's inequalities, and results like them, are often presented as
`no-go theorems' in the philosophical and physical literature. They
are portrayed as demonstrating, for instance, that no `locally causal'
\citep[][\textsection 7]{bell1990} description of the empirically
confirmed predictions of quantum mechanics is possible, that one
cannot ascribe `elements of reality', in EPR's sense
\citep[]{epr1935}, to quantum systems consistently with such
predictions, and so on. As I will argue below, however, thinking of
Bell's and related results in this way---without further
qualification---can be misleading, for they are not no-go theorems
\emph{per se}. Rather, they are best understood as establishing
certain general constraints on locally causal descriptions of joint
measurement outcomes. In order to turn Bell's and similar results into
no-go theorems, one must do more than just state them. One must also
interpret them in the context of the question one takes them to be
informing.

Associated with a given context are a number of additional
constraints, according to which we judge a locally causal description
to be either plausible or implausible in that context. Normally these
`plausibility constraints' are left unstated, and this is mostly
harmless most of the time, for the relevant context is usually
understood to be what I will below be calling the `theoretical'. Here
the question we take Bell's and related results to be answering is the
question of whether there is an alternative locally causal description
of the natural world that can recover the confirmed predictions of
quantum mechanics. The set of plausibility constraints on locally
causal descriptions---or anyway the general character of such
constraints---in the context of this question is implicitly understood
by all.

Bell's and related results are relevant to other contexts having to do
with physical systems as well, however. In particular, in what I will
below be calling the `practical' context, we are not concerned with
alternative theories of the natural world, classical or
otherwise. What we are rather concerned with are the kinds of physical
systems that we can build in order to recover a particular probability
distribution or set of measurement outcomes. As I will describe in
more detail below, the plausibility constraints we impose on locally
causal descriptions in this context are importantly different from
those we impose in the theoretical context. Thus the kinds of locally
causal description we think are worth considering as alternatives to
quantum description in the former context will be different from the
kinds of locally causal description we think are worth considering in
the latter.

This has interesting and important implications. \citet[]{cuffaroForthA}
has argued that distinguishing between these two contexts can help us
to understand the way certain physical resources are taken advantage
of in the science of quantum computation. The present article builds
upon Cuffaro's work.\footnote{To my knowledge, Cuffaro is the first
  to explicitly make this distinction. The general idea, however, of
  considering the difference between quantum and classical systems
  from the point of view of the practical context is implicit in the
  earlier work of \citet{pitowsky2002}.} In it I will argue that
making this distinction can help us to better understand the relative
power and scope of no-go theorems, and in so doing provide us with a
new point of view from which to consider the fundamental differences
between classical and quantum systems in general.

Consider, in particular, \citeauthor{ghz1989} (GHZ)'s so-called
`all-or-nothing' equality \citep{ghz1989}. It is typical in the
philosophical and foundational literature to see this equality
referred to as a more powerful refutation of local causality than
Bell's own statistical inequality. The reason for this is that, while
a violation of the GHZ equality can in principle be shown with a
single quantum experiment, a violation of Bell's requires repeated
quantum experiments to demonstrate (and even then only with
increasing, but never absolute,
confidence). \citet[p. 1131]{ghsz1990}, for example, write: ``This
incompatibility with quantum mechanics is stronger than the one
previously revealed for two-particle systems by Bell's inequality,
where no contradiction arises at the level of perfect correlations.''
\citet[p. 174]{clifton1991} likewise write, of their improvement on
the original GHZ proof, ``[o]ne difference between Bell's theorem and
ours is that his yields a statistical contradiction, whereas ours
leads to an algebraic one. So in one respect, our conclusion is
stronger.''\footnote{Note that, as this quote hints at,
  \citeauthor[]{clifton1991} do not argue that the GHZ proof is
  stronger than Bell's in every respect.} Mermin, for whom the
GHZ proof is ``spectacular'' \citeyearpar[p. 810]{mermin1993},
enthusiastically agrees: ``[t]his is an altogether more powerful
refutation of the existence of elements of reality than the one
provided by Bell's theorem ...''
\citeyearpar[p. 11]{mermin1990}. And Maudlin concurs: ``the GHZ scheme
brings home the problem for locality all the more sharply''
\citep[p. 26]{maudlin2011}. So, also, does Vaidman: ``The GHZ proof is
the most clear and persuasive proof of nonexistence of local hidden
variables'' \citeyearpar[p. 615]{vaidman1999}. Indeed, for Vaidman,
the significance of the GHZ proof is quite profound: ``analysis of the
GHZ work led me to accept the bizarre picture of quantum reality given
by the many worlds interpretation, ...'' (ibid., p. 616). Vaidman is
not alone in taking the GHZ equality as a starting point from which to
consider the question of the interpretation of quantum mechanics
\citep[e.g., see][]{hemmo2003}.

There is nothing inappropriate about this kind of talk, so long as it
is confined (as it presumably is for the above authors) to what I have
above called the theoretical context. When we move, however, from the
theoretical to the practical context, then interestingly, the above
statements are false. As I will elaborate upon in more detail below,
in the practical context, the all-or-nothing GHZ theorem loses its
force. In the practical context, in fact, it is only statistical
inequalities which can legitimately be turned into no-go theorems
(although, also interestingly, one must indeed go beyond Bell's
theorem, and its associated bipartite entangled states, to entangled
states of multipartite systems in order to show this). From this we
may say that from the `absolute' point of view---i.e. when considering
the limits of plausible classical physical description as such---it is
statistical inequalities, rather than the all-or-nothing GHZ, which
are more powerful in the sense alluded to by the authors above, for
statistical inequalities---and not the all-or-nothing GHZ---allow us
to see that there are quantum statistics that even we can't plausibly
build classical systems to reproduce.

The sections of this paper are the following: in \textsection
\ref{sec:nogo} I review CHSH's variant of Bell's inequality as well as
the GHZ equality. In \textsection \ref{sec:schemes} I describe some
schemes by which one could simulate, using a classical computer, some
of the statistics associated with Bell and GHZ states. In \textsection
\ref{sec:sims} the meaning of the term ``classical computer
simulation'' is reflected upon. The implications of these reflections
for our understanding of the relative strength of statistical no-go
theorems \emph{vis \'a vis} the all-or-nothing GHZ theorem are
considered in \textsection \ref{sec:compare}. Finally, their general
implications for our understanding of the limits of classical
description, and for our understanding of the fundamental aim of
physical investigation, are considered in \textsection
\ref{sec:gendisc}.

\section{`No-Go' Results}
\label{sec:nogo}

\subsection{The CHSH inequality}
\label{subsec:chsh}

Consider the singlet state, $| \Psi^- \rangle = 1/\sqrt 2(| \uparrow
\rangle| \downarrow \rangle - | \downarrow \rangle| \uparrow \rangle)$
of two spin-\sfrac{1}{2} particles. For a system in this state, the
expectation value for joint experiments on its two subsystems is given
by the following expression:
\begin{align}
\label{eqn:singlet}
\langle \sigma_m \otimes \sigma_n \rangle = - \hat{m} \cdot \hat{n} =
- \cos\theta.
\end{align}
Here $\sigma_m, \sigma_n$ represent spin-$m$ and spin-$n$ experiments
on the first (Alice's) and second (Bob's) subsystem, respectively,
with $\hat{m}, \hat{n}$ the unit vectors representing the orientations
of their two experimental devices, and $\theta$ the difference in
these orientations.

It is well known that it is not possible to provide an alternative
theory accounting for the predictions associated with this state if
that theory makes the very reasonable assumption that the
probabilities of local experiments on Alice's (and likewise Bob's)
subsystem are completely determined by her local experimental
setup together with a shared hidden variable $\lambda$ taken on by
both subsystems at the time the joint state is prepared (i.e. while
$A$ and $B$ are still physically interacting). For, given such a
theory, we will have that
\begin{align}
\label{eqn:factorise}
\langle \sigma_m \otimes \sigma_n \rangle_\lambda = \langle
A_\lambda(\hat{m}) \times B_\lambda(\hat{n}) \rangle,
\end{align}
where $A_\lambda(\hat{m}) \in \{\pm 1\}, B_\lambda(\hat{n}) \in \{\pm
1\}$ represent the results, given a specification of the hidden
variable $\lambda$, of spin experiments on Alice's and Bob's
subsystems. But consider the following expression relating the
expectation values of different combined spin experiments on Alice's
and Bob's subsystems for arbitrary directions $\hat{m}, \hat{m}',
\hat{n}, \hat{n}'$:
\begin{align*}
\mbox{\emph{CHSH}}:\quad | \langle \sigma_m \otimes \sigma_n \rangle +
\langle \sigma_m \otimes \sigma_{n'} \rangle | + | \langle \sigma_{m'}
\otimes \sigma_n \rangle - \langle \sigma_{m'} \otimes \sigma_{n'}
\rangle |.
\end{align*}
When \eqref{eqn:factorise} holds, one can show that
\begin{align}
\label{eqn:chsh}
\mbox{\emph{CHSH}} \leq 2.
\end{align}
The `CHSH inequality' \citep[][]{chsh1969} is one of a family of
similar expressions, known collectively as the `Bell
inequalities',\footnote{These are named after Bell, the discoverer of
  the first such result \citep[]{bell1964}.} which must be satisfied
by locally causal theories aiming to account for the statistics
associated with combined spin measurements.

The statistical predictions of quantum mechanics violate the CHSH
inequality for some experimental configurations. For example, let the
system be in the singlet state and let the unit vectors $\hat{m},
\hat{m}', \hat{n}, \hat{n}'$ (taken to lie in the same plane) have the
orientations $0, \pi/2, \pi/4, -\pi/4$ respectively. The differences,
$\theta$, between the different orientations (i.e., $\hat{m} -
\hat{n}, \hat{m} - \hat{n}', \hat{m}' - \hat{n}$, and $\hat{m}' -
\hat{n}'$) will all be in multiples of $\pi/4$ and we will have, from
\eqref{eqn:singlet}:
\begin{align*}
| \langle \sigma_m \otimes \sigma_n \rangle & + \langle \sigma_m \otimes
\sigma_{n'} \rangle | + | \langle \sigma_{m'} \otimes \sigma_n \rangle
- \langle \sigma_{m'} \otimes \sigma_{n'} \rangle | = 2\sqrt 2
\not\leq 2.
\end{align*}

We normally conclude from this that the predictions of quantum
mechanics for arbitrary orientations $\hat{m}$, $\hat{m}'$, $\hat{n}$,
$\hat{n}'$, cannot be reproduced by any alternative theory in which
all of the correlations between subsystems are due to a common
parameter endowed to them at state preparation. But note that these
predictions can be reproduced by such a hidden variables theory for
certain special cases. In particular, the inequality is satisfied by
quantum mechanics when $\hat{m}$ and $\hat{n}$, $\hat{m}$ and
$\hat{n}'$, $\hat{m}'$ and $\hat{n}$, and $\hat{m}'$ and $\hat{n}'$
are all oriented at angles with respect to one another that are given
in multiples of $\pi/2$. Note that these are the cases for which
Eq. \eqref{eqn:singlet} predicts perfect correlation ($\theta = \pi$),
perfect anti-correlation ($\theta = 0$), or no correlation at all
($\theta = \pi/2$).

For example, let $\lambda$ determine Alice's and Bob's measurement
results in the following way:
\begin{align}
\label{eqn:bellhvt}
A_\lambda(\hat{m}) & = \mbox{sign} (\hat{m} \cdot \hat{\lambda}),
\nonumber \\
B_\lambda(\hat{n}) & = - \mbox{sign} (\hat{n} \cdot \hat{\lambda}),
\end{align}
where $\mbox{sign}(x)$ is a function which returns the sign (+ or -)
of its argument. The reader can verify that \eqref{eqn:bellhvt} will
recover all of the statistical predictions associated with the singlet
state \eqref{eqn:singlet} just so long as the difference in
orientation between $\hat{m}$ and $\hat{n}$ is some multiple of
$\pi/2$ \citep[p. 16]{bell1964}.

\subsection{The GHZ equality}
\label{subsec:ghz}

We have just seen that the statistics associated with the singlet
state for measurement angles differing in proportion to $\pi/2$ are
reproducible in a local hidden variables theory such as
\eqref{eqn:bellhvt}. But this is not true for every entangled
state. In particular, for a system of three spin-\sfrac{1}{2}
particles in the state:
\begin{align}
\label{eqn:cat}
| \mbox{GHZ} \rangle = \frac{| \uparrow \rangle_a | \uparrow \rangle_b
  | \uparrow \rangle_c - | \downarrow \rangle_a | \downarrow \rangle_b
  | \downarrow \rangle_c}{\sqrt 2},
\end{align}
one can, by considering measurements of Pauli observables ($\sigma_x$,
$\sigma_y$, $\sigma_z$),\footnote{I have omitted the trivial
  identity operator $I$ from this list.} demonstrate a conflict
between the predictions of quantum mechanics and those of a suitably
local hidden variables theory. Note that the respective orientations
of different ends of an experimental apparatus set up to conduct an
experiment involving Pauli observables on a combined system will never
differ by anything other than an angle proportional to $\pi/2$.

To see how this conflict arises, note that in the GHZ
state,\footnote{The exposition below is based on Mermin's
  \citeyearpar[]{mermin1990}.} the eigenvalues associated with
$\sigma_x$ and $\sigma_y$ observables on individual subsystems are (as
always) $\pm 1$, while each of the tripartite observables,
\begin{align}
\label{eqn:ghzops}
\sigma_x^a \otimes \sigma_y^b  \otimes \sigma_y^c, \quad \sigma_y^a
\otimes \sigma_x^b \otimes \sigma_y^c, \quad \sigma_y^a \otimes
\sigma_y^b \otimes \sigma_x^c
\end{align}
(which, as the reader can verify, are compatible) takes the eigenvalue
$1$; i.e.,
\begin{align}
\label{eqn:ghzopseigs}
v(\sigma_x^a \otimes \sigma_y^b  \otimes \sigma_y^c) = v(\sigma_y^a
\otimes \sigma_x^b \otimes \sigma_y^c) = v(\sigma_y^a \otimes
\sigma_y^b \otimes \sigma_x^c) = 1.
\end{align}

Thus preparing a tripartite system in the state \eqref{eqn:cat} will
yield correlations, expressed by \eqref{eqn:ghzopseigs}, between the
results of certain measurements at the sites $a$, $b$, and $c$. Since
$a$, $b$, and $c$ may in general be quite distant from one another, it
is reasonable to assume, if one is reasoning classically, that the
subsystems measured at those sites became correlated while they were
still in physical interaction with one another (i.e., at state
preparation) by way of some shared common cause (represented by a
variable $\lambda$). Thus, at the time of measurement, nothing further
should influence an individual outcome aside from the local properties
of the experimental setup at a site and of the particle being measured
there. Thus the results of the combined measurements
\eqref{eqn:ghzopseigs} will be factorisable given $\lambda$; i.e.,

\begin{align}
\label{eqn:products}
v(\sigma_x^a \otimes \sigma_y^b  \otimes \sigma_y^c) = 1 =
v(\sigma_x^a) \times v(\sigma_y^b) \times v(\sigma_y^c), \nonumber \\
v(\sigma_y^a \otimes \sigma_x^b  \otimes \sigma_y^c) = 1 =
v(\sigma_y^a) \times v(\sigma_x^b) \times v(\sigma_y^c), \nonumber \\
v(\sigma_y^a \otimes \sigma_y^b  \otimes \sigma_x^c) = 1 =
v(\sigma_y^a) \times v(\sigma_y^b) \times v(\sigma_x^c).
\end{align}

But this cannot be. Multiplying the right hand sides of
\eqref{eqn:products}, and using the fact that $v(\sigma_y^a)^2 =
v(\sigma_y^b)^2 = v(\sigma_y^c)^2 = 1$, we have it that $v(\sigma_x^a)
\times v(\sigma_x^b) \times v(\sigma_x^c) = 1,$ which in turn (given
factorisability) implies that
\begin{align}
\label{eqn:contra1}
v(\sigma_x^a \otimes \sigma_x^b \otimes \sigma_x^c) = 1.
\end{align}
Quantum mechanically, however, if we take the product (we can since
they are compatible) of the observables in \eqref{eqn:ghzops}, then
since $\sigma_x\sigma_y = i\sigma_z = -\sigma_y\sigma_x$,
$\sigma_x\sigma_z = -i\sigma_y = -\sigma_z\sigma_x$, $\sigma_y\sigma_z
= i\sigma_x = - \sigma_z\sigma_y$, $\sigma_x\sigma_x =
\sigma_y\sigma_y = \sigma_z\sigma_z = I$, this must yield:
\begin{align}
\label{eqn:quantprod1}
& (\sigma_x^a \otimes \sigma_y^b \otimes \sigma_y^c)(\sigma_y^a
\otimes \sigma_x^b \otimes \sigma_y^c)(\sigma_y^a \otimes
\sigma_y^b \otimes \sigma_x^c) \nonumber \\
=\mbox{ } & -\sigma_x^a \otimes \sigma_x^b \otimes \sigma_x^c.
\end{align}
Since each of the observables in \eqref{eqn:ghzops} takes the
eigenvalue 1, this implies that
\begin{align}
\label{eqn:contra2}
v(\sigma_x^a \otimes \sigma_x^b \otimes \sigma_x^c) = -1.
\end{align}

Thus \eqref{eqn:contra1}, which we were led to through the assumption
of factorisability \eqref{eqn:products}, contradicts the (empirically
confirmed) predictions of quantum mechanics \eqref{eqn:contra2}. We
conclude, therefore, that the correlations expressed by
\eqref{eqn:ghzopseigs} do not admit of a locally causal description.

\section{Classically simulating quantum statistics}
\label{sec:schemes}

The results just described have unquestionably deepened our
understanding of the implications of our experience with quantum
phenomena. Yet the proper analysis and precise significance of these
implications for our understanding of nature has, unsurprisingly, been
hotly debated both by philosophers and physicists.\footnote{See, for
  instance, \citet[]{cushing1989}.} We will not engage directly in
those debates here. We will rather ask ourselves a different (but as
we will see later, not unrelated) question: (i) could one build a
classical machine to simulate the quantum measurement statistics
described above (and if so, how)? If the answer is yes, then (ii) are
there any quantum correlations that cannot be so simulated, or is it
the case that one can build classical systems to reproduce every
possible quantum mechanical correlational effect?

As it turns out, this is, in fact, an active area of research, and as
we will see shortly, we can indeed give an affirmative answer to
question (i). Later, in \textsection \ref{sec:sims}, we will consider
the philosophical significance of this. In particular we will see (in
\textsection \ref{sec:compare}) that it forces us to qualify some of
the conclusions we made in \textsection \ref{sec:nogo}. In the process
we will answer question (ii).

\subsection{GHZ statistics}
\label{subsec:ghzsim}

Consider, first, the case of GHZ correlations.\footnote{What follows
  is a summary of the model given in more detail in
  \citep{tessier2004, tessier2005}.} Table \ref{tab:ghzsim} depicts
a scheme for reproducing all of the Pauli measurement statistics
associated with the state \eqref{eqn:cat}.

\begin{table}
\begin{align*}
\begin{matrix}
\\
\hline
\sigma_x \\
\sigma_y \\
\sigma_z \\
I
\end{matrix}
\left|
\begin{matrix}
  a & b & c \\
  \hline
  -R_2R_3     & R_2     &  R_3 \\
  -iR_1R_2R_3 & iR_1R_2 &  iR_1R_3  \\
  R_1        & R_1     &  R_1 \\
  1          & 1       &  1
\end{matrix}
\right|
\begin{matrix}
\\
\hline
\mbox{ }\\
\mbox{ }\\
\mbox{ }\\
\mbox{ }.
\end{matrix}
\end{align*}
\caption{A scheme for reproducing all of the Pauli measurement
  statistics associated with the tripartite GHZ state.}
\label{tab:ghzsim}
\end{table}

In the table, $a$, $b$, and $c$ refer to the three subsystems of the
system. $\sigma_x$, $\sigma_y$, $\sigma_z$, and $I$ are the Pauli spin
operators. The shared variables $R_1$, $R_2$, and $R_3$ are values of
$\pm 1$ that are assigned to the various subsystems at state
preparation through some sequence of local interactions. It is assumed
that these variables possess determinate values prior to measurement
but that these values are completely `hidden' in the sense that they
can only be revealed by measurement, or what amounts to the same
thing: two identical state preparations will in general yield (with
equal likelihood) different values of $R_1$ (likewise for $R_2$ and
$R_3$).\footnote{Note that no such interpretation of $R_1$, $R_2$,
  and $R_3$ is given in either \citep{tessier2004} or
  \citep{tessier2005}, but it is implicit.}

To determine the outcome of a combined Pauli measurement on the
system, one multiplies the entries of the table corresponding to the
measurements performed on each subsystem, discarding any remaining
unsquared $i$'s. For example, a measurement of $\sigma_y$ on subsystem
$a$ is equivalent to $\sigma_y \otimes I \otimes I$. The result is
given by $-iR_1R_2R_3$, which, dropping the unsquared $i$, yields
$-R_1R_2R_3 = \pm 1$ with equal likelihood. Measuring $\sigma_y \otimes
\sigma_x \otimes \sigma_y$, on the other hand, will yield
\begin{align}
\label{eqn:context1}
v(\sigma_y \otimes \sigma_x \otimes \sigma_y) = -iR_1R_2R_3R_2iR_1R_3 =
-i^2R_1^2R_2^2R_3^2 = 1
\end{align}
with certainty. And so, on. It is easy to verify that Table
\ref{tab:ghzsim}'s predictions for every combined Pauli measurement
match up with those of quantum mechanics. And like those of quantum
mechanics, Table \ref{tab:ghzsim}'s predictions for combined Pauli
measurements on the GHZ state are not factorisable, since, for
example,
\begin{align}
\label{eqn:context2}
v(\sigma_y \otimes I \otimes I) \times v(I \otimes \sigma_x \otimes I)
\times v(I \otimes I \otimes \sigma_y) = -R_1R_2R_3R_2R_1R_3 = -1,
\end{align}
in contradiction with \eqref{eqn:context1}. The instances of $i$ in
the table are the analogue of quantum mechanical `nonlocal'
influences.

With only very minor tweaking, however, one can build a model similar
to the one represented in Table \ref{tab:ghzsim}, in which the
resulting correlations are both factorisable and consistent with all
of the predictions of quantum mechanics for these measurements. Our
tweak will be that we will allow the parties to classically signal to
one another as follows. Bob (measuring $b$) and Alice (measuring $a$)
will agree that he will send her a single classical bit indicating
whether or not he performed a $\sigma_y$ measurement on his
subsystem. Upon receipt of this bit, Alice should flip the sign of her
local outcome if either she or Bob (or if both of them) measured
$\sigma_y$. Thus instead of \eqref{eqn:context2} we will have:
\begin{align}
\label{eqn:context3}
v(\sigma_y \otimes I \otimes I) \times v(I \otimes \sigma_x \otimes I)
\times v(I \otimes I \otimes \sigma_y) = R_1R_2R_3R_2R_1R_3 = 1,
\end{align}
in agreement with \eqref{eqn:context1}. Note that this result is
consistent with each of the three measurements $\sigma_y \otimes I
\otimes I$, $I \otimes \sigma_x \otimes I$, $I \otimes I \otimes
\sigma_y$, for each of them separately will yield a value of
$\pm 1$ with equal probability.

This small addition---a single bit of classical
communication---suffices to make the outcome of every joint
measurement of Pauli observables factorisable in the model. The scheme
generalises. For an $n$-partite system in the GHZ state $1/\sqrt
2(\mbox{$| \uparrow \rangle$}^{\otimes n} \pm | \downarrow
\rangle^{\otimes n})$, only $n-2$ bits of classical communication are
required. The same is true, moreover, not only for $n$-partite GHZ
states, but also for any $n$-partite state in which each superposition
term is expressible as a product of Pauli eigenstates (details are
given in \citealt{tessier2004}).

One can imagine building a classical computational system, made up of
smaller spatially separated computational subsystems, utilising a
small amount of classical communication as described above, to
instantiate the description given in Table \ref{tab:ghzsim}. This is
the sense in which one may say that the correlations between the
results of Pauli measurements on systems in the GHZ state are
`efficiently classically simulable'. I will elaborate upon this in
more detail in \textsection\ref{sec:sims}. But let us first consider
how to simulate measurement statistics on a system in the singlet
state.

\subsection{Singlet statistics}
\label{subsec:singletsim}

As we saw earlier \eqref{eqn:bellhvt}, Bell himself provided a local
hidden variables description to account for Pauli measurement
statistics in the singlet state.\footnote{The theory given in
  \eqref{eqn:bellhvt} applies not just to Pauli measurements, of
  course, but more generally, as we saw, to any combined spin
  measurement in which the angles at the respective ends of the
  experimental apparatus differ in proportion to $\pi/2$.} Presumably,
we could build a classical computer to instantiate that
description. Thus Pauli measurements on the singlet state are
efficiently classically simulable in this sense. In this case no
communication is needed, in fact. This is evident from
\eqref{eqn:bellhvt}. It also follows from the fact that Pauli
measurement statistics on any $n$-partite state, whose superposition
terms are expressible as products of Pauli eigenstates, can be
simulated using $n-2$ bits. The singlet state is one such state for
which $n = 2$. Thus for the singlet state, Pauli measurements can be
simulated using $2 - 2 = 0$ bits of communication.

It turns out that for the singlet state, we can simulate far more than
just Pauli measurements if we allow ourselves only a single additional
bit. Astoundingly, we can actually recover the statistics associated
with arbitrary projective measurements by using the following method
\citep{toner2003}. Randomly choose, independently, two unit vectors
$\hat{\lambda}_1$ and $\hat{\lambda}_2$. At state preparation, share
them with Alice and Bob, and instruct them to take their particles
with them to separate distant locations. Once there, have Alice
measure her particle along the direction $\hat{m}$, and output the
result $A = -\mbox{sign}(\hat{m} \cdot \hat{\lambda}_1).$ Have her
then send a single classical bit $c = \mbox{sign}(\hat{m}
\cdot\hat{\lambda}_1) \times \mbox{sign}(\hat{m} \cdot\hat{\lambda}_2)
= \pm 1$ to Bob, who finally measures along $\hat{n}$ and outputs the
result $B = \mbox{sign}[\hat{n} \cdot (\hat{\lambda}_1 +
  c\hat{\lambda}_2)].$

\section{What is a classical computer simulation?}
\label{sec:sims}

We have just seen how to, using only a few additional resources,
classically simulate the quantum measurement statistics described in
\textsection \ref{sec:nogo}. We saw, specifically, that only a single
bit is needed to recover the statistics associated with arbitrary
projective measurements on the singlet state, and that only $n-2$ bits
are needed to recover the statistics associated with Pauli
measurements on $n$-partite GHZ states. Let us now reflect on this. To
begin with, let us try and explicate somewhat more precisely what is
meant by the statement that a particular set of quantum correlational
statistics has been `classically simulated'.

What is a classical computer simulation, then? Most obviously,
perhaps, we may say, to start with, that a classical computer
simulation is something performable by a classical computer. Let us
consider what is meant by the latter. A classical computer, whatever
else it is, is a classical physical system, i.e., a system which can
be given a classical physical description. Some of the characteristics
of classical description are the following. First, a complete
classical description of a system, which in general will consist of
many subsystems, is always separable into complete descriptions of
those individual subsystems. Second, a classical description of the
interactions between a system's subsystems will not violate classical
physical law. For instance, the speed by which such interactions
propagate should be less than or equal to the speed of
light.\footnote{I am taking `classical' in the sense in which it is
  typically used in discussions of quantum mechanics; i.e. I take
  classical theory to include special and general relativity.} Third,
and importantly, the behaviour of classical systems is always
describable in principle in terms of local causes and effects, in the
metaphysically deflationary sense explicated by
\citet[]{bell1990}. Thus, while a classical system might evince
certain correlations between experimental results on its various
subsystems---which, in general, can be quite distant from one
another---since the system is classical, these correlations will
always be describable as arising from a common source situated in the
intersection of the prior light cones of the associated measurement
events. These correlations, once we take into account their common
causes, are always, that is, factorisable. Classical systems do not
violate Bell (or GHZ, or similar) inequalities. Classical computer
simulations, therefore, which are run on classical computers, which
are instantiated by classical physical systems, do not either.

This is perfectly obvious. And yet it might still strike one as
strange. The results of, say, a Bell experiment seem to be evidence
for some sort of influence (even if only benign) between the spatially
separated subsystems of entangled systems, and we usually take this to
demonstrate that quantum systems are `nonlocal' (or perhaps
`nonseparable') in some sense (see, e.g., \citealt{maudlin2011});
i.e., we take such influences to demonstrate that no locally causal
description---no `local hidden variables theory'---of the outcomes of
these experiments is possible. But the influence of one spatially
separated subsystem on another is precisely what we find in the
protocols designed to reproduce these outcomes described in
\textsection \ref{sec:schemes}. And yet these are locally causal
descriptions. How can this be?

A way of resolving this tension has been convincingly provided by
\citet[]{cuffaroForthA}. I will not reproduce that discussion in its
entirety here, but rather only summarise the main points emerging from
it that are relevant to our own investigation (for a full elaboration
and defense of these points, see the aforementioned article). Cuffaro
argues that when we make a judgement to the effect that no local
hidden variables theory is capable of recovering the quantum
mechanical predictions associated with a Bell experiment, implicit in
this judgement is a set of constraints upon the kinds of local hidden
variables theory that we are willing to entertain. To put it another
way: with a little imagination it is always possible to conceive of
loopholes to the Bell inequalities, but only some of these
loopholes---the plausible ones---are worth the bother of trying to
close. This means, however, that a conclusion such as ``no (plausible)
hidden variables theory of type $X$ is able to reproduce the
experimental predictions of quantum mechanics with respect to
preparation procedure $Y$'' cannot really be forced on the basis of
Bell's (or a similar) (in)equality alone. Strictly speaking, to force
such a conclusion, what one means by `plausible' must be spelled out
first.

Usually, of course, we do not really need to be explicit. There is a
set of criteria by which we mark a hidden variables theory as
plausible which is implicitly understood by everyone. This set
includes such things as consistency with our other theories of
physics, with the body of our experiential knowledge in general; we
expect a candidate local hidden variables theory not to be
conspiratorial, and so on. These and other plausibility constraints
are associated with what Cuffaro calls the `theoretical' context,
wherein we take Bell's and related results to enlighten questions such
as `is there an alternative (locally causal) description of the
natural world that can recover the confirmed predictions of quantum
mechanics?'

There are other contexts, besides the theoretical, to which Bell's and
related (in)equalities are applicable. In particular, Cuffaro
describes the `practical' context, wherein our concern is not with
alternative theories of the natural world, but rather with what we are
capable of building with the aim of (classically) reproducing the
statistical predictions of quantum mechanics. In the practical context
we are concerned, in other words, not with `local hidden variables'
descriptions of the natural world, but with `practical classical'
descriptions of machines that are possible for us to build to achieve
particular ends (in the sequel I will be using, with Cuffaro, the
generic term `locally causal description' to refer to both).

Now consider the two classical schemes for simulating quantum
statistics described in \textsection \ref{sec:schemes}. Either of
these can be thought of as an alternative locally causal description
of the observation of a given set of measurement statistics in the
following way. Begin by asking what we really need to recover in an
alternative locally causal description of a combined measurement
outcome. The answer invariably will be: just the actual observation of
the combined outcome itself. What I mean is this: although quantum
mechanics posits its own particular dynamics---involving, for example,
the state collapse of a subsystem of a Bell pair upon the measurement
of the other subsystem---an alternative description of the events
leading up to the observation of a combined measurement outcome can in
principle disagree entirely with the quantum mechanical dynamical
description. All that an alternative locally causal description needs
to be faithful to in principle is our actual observation of the
combined result.

\begin{figure}
\begin{center}
  \includegraphics[scale=0.25]{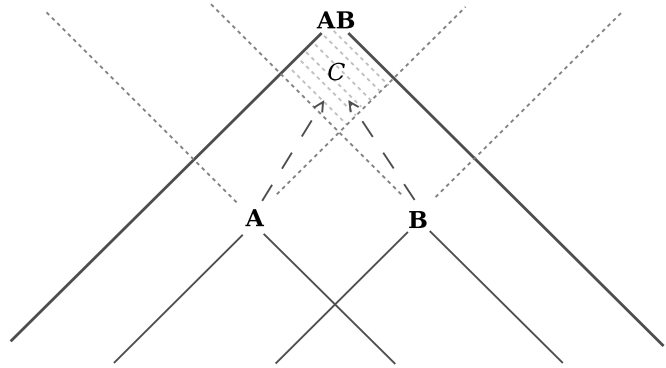}
\end{center}
\caption{A conceivable explanation of the observation of the result of
  a joint experimental outcome $\mathbf{AB}$ which posits local
  (conspiratorial) influences interacting within region $C$. This
  figure is adapted from \citet[]{cuffaroForthA}.}
\label{fig:localVnonlocal}
\end{figure}

Importantly, in order to be in a position to assert that one has
observed a combined measurement outcome, one must first have combined
the individual measurement outcomes; Alice and Bob must report their
results to Candice (or to each other) over tea or by telephone or in
writing or by using some other physical means. This is absolutely
necessary and there simply is no escaping it. While these results are
being brought together, however, Alice and Bob have time to
(conspiratorially) exchange---at subluminal velocity---a series of
finite signals with one another, and thus coordinate---and if
necessary, `correct'---their individual measurement outcomes in the
manner described in \textsection \ref{sec:schemes}. But notice now
that with respect to the observation of the combined measurement
outcome---which, to reiterate, is ultimately what we are required to
explain---all of this conspiratorial signalling will have occurred in
its past light cone. It is thus part of a posited state
preparation---a common cause---for that combined outcome and it thus
comprises part of a classical, locally causal, alternative description
of that combined outcome (see figure \ref{fig:localVnonlocal}).

In the theoretical context, of course, such a locally causal
description would be ruled out as wildly implausible (although see
Kent's \citeyearpar[]{kent2005} `Collapse Locality Loophole' for an
account that is essentially similar). It would simply be, for this
author at any rate, too conspiratorial a loophole to take seriously,
let alone to spend any of one's research funds to try and close. But
in the practical context, Cuffaro convincingly argues that we will not
fret if our alternative characterisation of a set of measurement
statistics is conspiratorial or overly \emph{ad hoc} from the
theoretical point of view, for in the practical context it is always
assumed that the system under consideration has been purposely
designed to recover those statistics. What we will rule out as
implausible, rather, are only descriptions of classical systems that
would be too `hard' for someone to build, where `hard' and likewise
`easy' are defined naturalistically---i.e. by appealing to our best
scientific theory of the subject: computational complexity theory.

To illustrate: imagine someone, call him Ray, attempting to convince
you that a large black box sitting in a hangar at the airport is a
quantum computer. Suppose that he either cannot or will not open the
box, but that he rather attempts to convince you of this by having the
computer run a calculation whose result depends (according to him) on
the prior generation of a GHZ state. Now if you are skeptical of this
claim, it will then be incumbent upon you to produce an alternative
classical description---a blueprint, say---of the goings on within the
box which explain your observations in the hangar. If, however, the
best that you can do is specify a classical model for your
observations that requires an infinite---or anyway some
`enormous'---number of additional resources as compared to a quantum
model, then in that case Ray may legitimately dismiss your objection,
for it is implausible, he will say, that he or anyone could have built
a machine conforming to your description. On the other hand if you can
specify a machine whose description requires only `a few' additional
resources, in the manner of \textsection \ref{sec:schemes}, then your
skepticism is justified unless Ray provides you with additional
evidence for his claim.\footnote{This last qualification is
  important. Presumably if the box were open, for instance, there
  would be ways to test for and close your loophole, at least in
  this simple scenario. This does not affect the conceptual point
  being made here, however.}

Now in complexity theoretic terms, to solve a problem using `an
enormous' number of resources means that these resources are
exponential with respect to the size of the input to the problem, $n$;
i.e. on the order of $k^n$. `A few' resources, in contrast, are at
most polynomial with respect to $n$; i.e., on the order of
$n^k$.\footnote{For a good and thorough text on the subject   of
  computational complexity, consult \citet{papadim1994}.} The
protocols described in \textsection \ref{sec:schemes} require only a
few additional resources to implement in this sense. In particular
they require a number of additional resources that is only linear in
$n$. For this reason, they and other efficient simulations of quantum
measurement statistics may be thought of as alternative locally causal
descriptions of those measurement statistics that are plausible in the
practical context.\footnote{It must not be forgotten, of course,
  that at least in the case of GHZ states, it is only the outcomes
  associated with Pauli measurements that are efficiently
  classically simulable in the sense just described. We will return
  to this point shortly.}

\section{Comparing the all-or-nothing GHZ with statistical
  (in)equalities}
\label{sec:compare}

We saw in \textsection \ref{subsec:singletsim} that if we would
like to classically simulate the measurement statistics associated
with Pauli measurements on a system in the singlet state, we can do so
without the aid of any classical communication. To recover the
statistics for arbitrary measurements, on the other hand, we require a
single classical bit propagated at subluminal speed. In \textsection
\ref{subsec:ghzsim} we saw that recovering the statistics associated
with Pauli measurements on a system in a GHZ state requires only a
number of additional resources that is linear in the number of
systems, $n$. All of these measurement statistics are plausibly
recoverable, therefore, in the practical context.

What if we would like to recover the statistics associated with
arbitrary measurements on a system in a GHZ state? Unfortunately,
unlike the singlet and other Bell states, it does not seem likely that
we could plausibly do so. \citet[p. 116]{tessier2004} notes that the
number of classical bits required to model the quantum mechanical
predictions associated with arbitrary projective measurements on
$n$-partite states (for $n \geq 3$) in a model like the one pictured
in Table \ref{tab:ghzsim} is likely to be unbounded. This is
consistent with other results due to \citet[]{jozsa2003}, and to
\citet[]{abbott2010}, who show (using different methods) that for
systems in pure states, an exponential speed up of quantum over
classical computation is possible only when one has available
multipartite entanglement with a number of systems $n \geq 3$.

Returning now to the GHZ argument which we discussed in \textsection
\ref{subsec:ghz}. As is evident from some of the quotations we
surveyed in \textsection \ref{sec:intro}, it is hard to overestimate
the impact that the GHZ proof has had upon the physical and
philosophical communities. To quote Mermin at length:

\begin{quote}
This is an altogether more powerful refutation of the existence of
elements of reality than the one provided by Bell's theorem for the
two-particle EPR experiment. Bell showed that the elements of reality
inferred from one group of measurements are incompatible with the
\emph{statistics} produced by a second group of measurements. Such a
refutation cannot be accomplished in a single run, but is built up
with increasing confidence as the number of runs increases [...] In
the GHZ experiment, on the other hand, the elements of reality require
a class of outcomes to occur \emph{all} of the time, while quantum
mechanics \emph{never} allows them to occur
\citep[p. 11]{mermin1990}.
\end{quote}

As we saw earlier, however, before one can properly make statements
like this one concerning the relative power of the Bell and GHZ
(in)equalities, one must first be clear on the context in which such a
statement is being made. From a theoretical point of view it may be
that the GHZ argument is more powerful than Bell's in the above
sense. Thus I take no issue with Mermin's statement as long as it is
qualified in this way, as I assume it implicitly is for Mermin. This
said, in the practical context, Mermin's statement is actually
false. For from a practical point of view, both the Bell and GHZ
(in)equalities are equally weak, in the sense that plausible
alternative locally causal descriptions can be provided to account for
the associated statistics in both cases.

The contrast being drawn in the quote above, however, and in the
statements of the other philosophers and physicists cited in
\textsection \ref{sec:intro}, is not specifically between the results
of Bell and GHZ. It is rather between the GHZ equality insofar as it
is an all-or-nothing result, and Bell's inequality insofar as it is a
statistical result. But as we have seen, it is quite easy to recover
the statistics associated with Pauli measurements on the GHZ
state---the measurements for which an all-or-nothing equality can be
proved---in the practical context. In order to use the GHZ state as
the basis for a \emph{bona fide} no-go theorem in this context, one
must consider measurements outside of the Pauli group, for the
statistics associated with these measurements cannot be plausibly
recovered in a locally causal model in this context. Once one does so,
however, the ensuing correlations between individual measurement
results will no longer be strict but probabilistic. Thus it is
statistical results, rather than the all-or-nothing GHZ, which are
more powerful in the practical context.

\section{General Discussion}
\label{sec:gendisc}

We have chosen to call classical computer simulations of quantum
mechanical correlational statistics locally causal alternative
descriptions of those correlational statistics. This is not the way it
is normally put in the quantum information literature. What is
normally claimed for these schemes is rather that they quantify---in
terms of the number of classical bits required to reproduce them---the
departure from classicality of quantum correlations \citep[see,
  e.g.,][]{rosset2013}.

While this way of framing their significance is certainly not
incorrect (and it is certainly, moreover, useful), from a more
philosophical point of view it is preferable to, so to speak, call a
spade a spade. As the example above involving the black box in the
hangar makes clear, a plausible description of a locally causal
computer simulation of a class of quantum phenomena is of the same
general kind as a plausible locally causal theoretical description of
a class of quantum phenomena. Both kinds of description represent
alternative plausible stories one can put forward if one is skeptical
regarding the quantum mechanical description of some set of
observations. The essential difference between the theoretical and
practical contexts lies in the different presuppositions we make
concerning the origin of the physical systems under consideration in
each case. In the practical context we presuppose that a particular
system has been designed and built by some rational agent for a
particular purpose. We do not presuppose nature to be purposeful in
this way in the theoretical context.\footnote{I am not expressing a
  theological opinion with this statement. I mean only that we do
  not presuppose this---not in this century, at any rate---for the
  purposes of scientific investigation.}

Additionally, calling a spade a spade helps us to understand one
reason why the practical context is---or anyway should
be---interesting from a philosopher of physics' point of view. Indeed,
the reader may have been wondering why she should care at all about
the practical context---why a philosopher of physics should care for
anything but the correct theory of the natural world and for its
deeper consequences for our fundamental understanding of the ontology
of that world. To help appreciate why we should care, imagine if it
were possible to easily build classical machines to reproduce the
statistics associated with every quantum mechanical measurement,
regardless of the state and the number of systems such a measurement
were performed upon. If this were possible it would surely be of
immense philosophical interest, for it would signify that we can build
classical physical systems to reproduce every observable quantum
mechanical correlational effect! It is hard to imagine this not having
some (indirect) bearing on our interpretation of the quantum
formalism, or at least on our general metaphysical world view.

I will not speculate as to what that would be, for as we saw in the
previous section, it turns out that building such universal simulators
is not plausible. And yet it is still quite interesting
philosophically to know that it is both possible---and
plausible---that we could build classical physical systems to fully
reproduce the observable correlational behaviour of systems in Bell
states. It is also interesting to know that there are quantum
correlational phenomena associated with other states that we cannot
plausibly build classical machines to reproduce. The reason all of
this is interesting is that in both the practical and theoretical
contexts we are speaking about physical systems, after
all. Considering the practical context can thus help us to understand
the very limits of classical description. It can help us to
understand, that is, just how far quantum physical description as such
outstrips classical physical description as such---whether there are
quantum statistics that even we couldn't plausibly build classical
systems to reproduce.

On the other hand, noting that there are important ways in which the
presuppositions appropriate to the practical and theoretical contexts
differ encourages us to be wary of conflating these contexts
inappropriately. Jeffrey Bub's claim \citeyearpar{bub2004}, for
instance, that the fundamental aim of physics consists in the
representation and manipulation of information, seems to me be in
danger of doing so. Bub's argument for this conclusion is based on the
fact that, within the abstract $C^*$ algebraic framework, one can
characterise quantum theory purely in terms of a set of information
theoretic constraints (specifically: no signalling, no broadcasting,
and no unconditionally secure bit commitment),\footnote{See
  \citet[]{myrvold2010}, however, for some reasons to be skeptical
  regarding the significance claimed for this characterisation
  result.} and that given these constraints, any alternative
mechanical theory which aims to solve the measurement problem faces an
in principle problem of underdetermination. This means, for Bub, that
``... our measuring instruments ultimately remain black boxes at some
level.'' \citep[p. 243]{bub2004}.

It will take us too far distant from the primary concern of this paper
to consider Bub's arguments for this last claim in detail, but let me
just say here that even if this claim---which, for different reasons,
I am generally sympathetic towards, and which, to be fair, actually
constitutes the main thrust of Bub's paper---is true,\footnote{For a
  criticism, see \citet[\textsection 8.3]{timpson2013}.} it does
not obviously follow from it that ``The appropriate aim of physics at
the fundamental level [is] the representation and manipulation of
information'' \citep[p. 242]{bub2004}, or that ``An entangled state
should be thought of as a nonclassical communication channel that we
have discovered to exist in our quantum universe, i.e., as a new sort
of nonclassical `wire''' \citep[p. 262]{bub2004}. All that would seem,
\emph{prima facie}, to follow from the claim that our mechanical
description of nature must run short is that we cannot descend below
the level of the language of information theory when characterising
quantum phenomena. But one could argue that the proper lesson to draw
from this is merely that, as a result, we must be all the more careful
to be precise about what it is we are speaking of when we are using
informational language---careful to distinguish, that is, the
circumstances in which we are discussing the ways in which we can use
physical systems, from the circumstances in which we are describing
the nature of the physical world itself as it exists in itself
(i.e. without our intervention).

The idea that we can and do make such distinctions lies at the very
basis of our foregoing discussion. Our analysis in \textsection
\ref{sec:sims} began from the fact that our conceptions of `classical
system buildable by a human being' and of `classical system existing
naturally' are---at least pre-theoretically---very different
conceptions. We then explicated these pre-theoretic conceptions more
precisely in terms of differing sets of plausibility constraints
associated with each. What would seem to be asserted by one
sympathetic to the view expressed above, however, is that our
pre-theoretic intuitions are false---that really, after all, there is
only one context worth considering: what we have above referred to as
the practical context.

It is not incoherent to hold such a view. But if it really is the case
that quantum theory is ultimately about nothing other than
representing and processing information---that (as seems to be implied
by this) the practical context should be our primary concern \emph{vis
  \'a vis} foundational investigation---then it is hard to see
why GHZ correlations, in particular, should surprise anyone. For from
the practical point of view, GHZ correlations are no more inexplicable
than the correlations between the colours of Professor Bertlemann's
socks (if you happen to notice that one of these is pink, you can rest
assured that the other is not; see: \citealt[]{bell1981}). In other
words, a plausible locally causal explanation, in the sense of
\textsection \ref{sec:schemes}, is readily available for both of these
phenomena in the practical context. But in fact we are surprised by
GHZ correlations (recall our survey of the reactions to the equality
in \textsection \ref{sec:intro}), and it is incumbent upon one
sympathetic to a view like Bub's to explain to us why this is so.

Let me close this section by noting that the foregoing considerations
are intended, not so much as an argument against a view like
Bub's. Rather, the sketch of an argument contained in the foregoing
few paragraphs should be regarded as an invitation, to anyone
sympathetic to the view that quantum theory just is about representing
and manipulating information---a much stronger and deeper claim than
the statement that the description of our measuring instruments must
necessarily remain opaque at some level---to clarify this position in
light of the points that have been made in this paper.

\section{Conclusion}
\label{sec:summ}

We reviewed the Bell and GHZ (in)equalities in \textsection
\ref{sec:nogo}, and then in \textsection \ref{sec:schemes} we saw
schemes by which one could simulate, using a classical computer, the
statistics associated with those (in)equalities. In \textsection
\ref{sec:sims} I argued that classical computer simulations are
locally causal descriptions in Bell's sense and hence do not violate
the Bell or GHZ (in)equalities, and I argued that this fact is not in
tension with our normal judgements regarding the significance of these
(in)equalities, as long as one understands that our normal judgements
are not valid in the context of a discussion of machines designed to
achieve a particular purpose. I then argued, in \textsection
\ref{sec:compare}, that this has implications for our understanding of
the relative strength of the GHZ all-or-nothing equality
\emph{vis-\'a-vis} statistical inequalities, and I made the perhaps
surprising observation that the former actually has no force in the
practical context, despite being a quite remarkable result in the
theoretical context. I ended, in \textsection \ref{sec:gendisc}, by
discussing the general implications of all of the foregoing for our
understanding of the limits of classical description, and for our
understanding of the fundamental aim of physical investigation.

Both in the practical and in the theoretical contexts, one's concern
is ultimately a very concrete one; i.e., one is in both cases
concerned with plausible descriptions of actual physical systems
existing in the world. The difference between the two contexts enters
essentially in the different presuppositions we make concerning the
origin of the physical systems under consideration. In the practical
context we presuppose that a particular system has been designed and
built by some rational agent for a particular purpose. We do not
presuppose this in the theoretical context. But because both contexts
are concerned with describing physical systems, both contexts are
relevant to our understanding of physical systems. And the practical
context is particularly illuminating in that it allows us to
investigate more thoroughly the possibilities inherent in physical
systems. It helps us to probe further and to understand better just
what the limits of classical description are, and just how far quantum
description outstrips it. At the same time we must remain on guard not
to be misled into thinking that the practical context can supersede
the theoretical context for the purposes of our foundational and
philosophical investigations into the nature of the (quantum) world.

\bibliographystyle{apa-good}
\bibliography{Bibliography}{}

\begin{thebibliography}{27}
\expandafter\ifx\csname natexlab\endcsname\relax\def\natexlab#1{#1}\fi
\expandafter\ifx\csname url\endcsname\relax
  \def\url#1{{\tt #1}}\fi
\expandafter\ifx\csname urlprefix\endcsname\relax\def\urlprefix{URL }\fi

\bibitem[{Abbott(2012)}]{abbott2010}
Abbott, A.~A. (2012).
\newblock The {Deutsch-Jozsa} problem: De-quantisation and entanglement.
\newblock {\em Natural Computing\/}, {\em 11\/}, 3--11.

\bibitem[{Bell(2004 {[1964]})}]{bell1964}
Bell, J.~S. (2004 {[1964]}).
\newblock On the {Einstein-Podolsky-Rosen} paradox.
\newblock In {\em Speakable and Unspeakable in Quantum Mechanics\/}, (pp.
  14--21). Cambridge: Cambridge University Press.

\bibitem[{Bell(2004 {[1981]})}]{bell1981}
Bell, J.~S. (2004 {[1981]}).
\newblock Bertlmann's socks and the nature of reality.
\newblock In {\em Speakable and Unspeakable in Quantum Mechanics\/}, (pp.
  139--158). Cambridge: Cambridge University Press.

\bibitem[{Bell(2004 {[1990]})}]{bell1990}
Bell, J.~S. (2004 {[1990]}).
\newblock La nouvelle cuisine.
\newblock In {\em Speakable and Unspeakable in Quantum Mechanics\/}, (pp.
  232--248). Cambridge: Cambridge University Press.

\bibitem[{Bub(2004)}]{bub2004}
Bub, J. (2004).
\newblock Why the quantum?
\newblock {\em Studies in History and Philosophy of Modern Physics\/}, {\em
  35\/}, 241--266.

\bibitem[{Clauser et~al.(1969)Clauser, Horne, Shimony, \& Holt}]{chsh1969}
Clauser, J.~F., Horne, M.~A., Shimony, A., \& Holt, R.~A. (1969).
\newblock Proposed experiment to test local hidden-variable theories.
\newblock {\em Physical Review Letters\/}, {\em 23\/}, 880--884.

\bibitem[{Clifton et~al.(1991)Clifton, Redhead, \& Butterfield}]{clifton1991}
Clifton, R.~K., Redhead, M. L.~G., \& Butterfield, J.~N. (1991).
\newblock Generalization of the {Greenberger-Horne-Zeilinger} algebraic proof
  of nonlocality.
\newblock {\em Foundations of Physics\/}, {\em 21\/}, 149--184.

\bibitem[{Cuffaro(forthcoming)}]{cuffaroForthA}
Cuffaro, M.~E. (forthcoming).
\newblock On the significance of the {G}ottesman-{K}nill theorem.
\newblock {\em The British Journal for the Philosophy of Science\/}.
\newblock Advance access version available at:
  \href{http://doi.org/10.1093/bjps/axv016}
  {http://doi.org/10.1093/bjps/axv016}.

\bibitem[{Cushing \& McMullin(1989)}]{cushing1989}
Cushing, J.~T., \& McMullin, E. (Eds.)  (1989).
\newblock {\em Philoophical Consequences of Quantum Theory\/}.
\newblock Notre Dame: University of Notre Dame Press.

\bibitem[{Einstein et~al.(1935)Einstein, Podolsky, \& Rosen}]{epr1935}
Einstein, A., Podolsky, B., \& Rosen, N. (1935).
\newblock Can quantum-mechanical description of physical reality be considered
  complete?
\newblock {\em Physical Review\/}, {\em 47\/}, 777--780.

\bibitem[{Greenberger et~al.(1990)Greenberger, Horne, Shimony, \&
  Zeilinger}]{ghsz1990}
Greenberger, D.~M., Horne, M.~A., Shimony, A., \& Zeilinger, A. (1990).
\newblock Bell's theorem without inequalities.
\newblock {\em American Journal of Physics\/}, {\em 58\/}, 1131--1143.

\bibitem[{Greenberger et~al.(1989)Greenberger, Horne, \& Zeilinger}]{ghz1989}
Greenberger, D.~M., Horne, M.~A., \& Zeilinger, A. (1989).
\newblock Going beyond {B}ell's theorem.
\newblock In M.~Kafatos (Ed.) {\em Bell's Theorem, Quantum Theory and
  Conceptions of the Universe\/}, (pp. 69--72). Dordrecht: Kluwer Academic
  Publishers.

\bibitem[{Hemmo \& Pitowsky(2003)}]{hemmo2003}
Hemmo, M., \& Pitowsky, I. (2003).
\newblock Probability and nonlocality in many minds interpretations of quantum
  mechanics.
\newblock {\em The British Journal for the Philosophy of Science\/}, {\em
  54\/}, 225--243.

\bibitem[{Jozsa \& Linden(2003)}]{jozsa2003}
Jozsa, R., \& Linden, N. (2003).
\newblock On the role of entanglement in quantum-computational speed-up.
\newblock {\em Proceedings of the Royal Society of London. Series A.
  Mathematical, Physical and Engineering Sciences\/}, {\em 459\/}, 2011--2032.

\bibitem[{Kent(2005)}]{kent2005}
Kent, A. (2005).
\newblock Causal quantum theory and the collapse locality loophole.
\newblock {\em Physical Review A\/}, {\em 72\/}, 012107.

\bibitem[{Maudlin(2011)}]{maudlin2011}
Maudlin, T. (2011).
\newblock {\em Quantum Non-Locality and Relativity\/}.
\newblock Cambridge, MA: Wiley-Blackwell, third ed.

\bibitem[{Mermin(1990)}]{mermin1990}
Mermin, N.~D. (1990).
\newblock What's wrong with these elements of reality?
\newblock {\em Physics Today\/}, {\em 43\/}, 9--11.

\bibitem[{Mermin(1993)}]{mermin1993}
Mermin, N.~D. (1993).
\newblock Hidden variables and the two theorems of {J}ohn {B}ell.
\newblock {\em Reviews of Modern Physics\/}, {\em 65\/}, 803--815.

\bibitem[{Myrvold(2010)}]{myrvold2010}
Myrvold, W.~C. (2010).
\newblock From physics to information theory and back.
\newblock In A.~Bokulich, \& G.~Jaeger (Eds.) {\em Philosophy of Quantum
  Information and Entanglement\/}, (pp. 181--207). Cambridge: Cambridge
  University Press.

\bibitem[{Papadimitriou(1994)}]{papadim1994}
Papadimitriou, C.~H. (1994).
\newblock {\em Computational Complexity\/}.
\newblock New York: Addison-Wesley.

\bibitem[{Pitowsky(2002)}]{pitowsky2002}
Pitowsky, I. (2002).
\newblock Quantum speed-up of computations.
\newblock {\em Philosophy of Science\/}, {\em 69\/}, S168--S177.

\bibitem[{Rosset et~al.(2013)Rosset, Branciard, Gisin, \& Liang}]{rosset2013}
Rosset, D., Branciard, C., Gisin, N., \& Liang, Y.-C. (2013).
\newblock Entangled states cannot be classically simulated in generalized
  {B}ell experiments with quantum inputs.
\newblock {\em New Journal of Physics\/}, {\em 15\/}, 053025.

\bibitem[{Tessier(2004)}]{tessier2004}
Tessier, T.~E. (2004).
\newblock {\em Complementarity and Entanglement in Quantum Information
  Theory\/}.
\newblock Ph.D. thesis, The University of New Mexico, Albuquerque, New Mexico.

\bibitem[{Tessier et~al.(2005)Tessier, Caves, Deutsch, \& Eastin}]{tessier2005}
Tessier, T.~E., Caves, C.~M., Deutsch, I.~H., \& Eastin, B. (2005).
\newblock Optimal classical-communication-assisted local model of $n$-qubit
  {Greenberger-Horne-Zeilinger} correlations.
\newblock {\em Physical Review A\/}, {\em 72\/}, 032305.

\bibitem[{Timpson(2013)}]{timpson2013}
Timpson, C.~G. (2013).
\newblock {\em Quantum Information Theory \& the Foundations of Quantum
  Mechanics\/}.
\newblock Oxford: Oxford University Press.

\bibitem[{Toner \& Bacon(2003)}]{toner2003}
Toner, B.~F., \& Bacon, D. (2003).
\newblock Communication cost of simulating {B}ell correlations.
\newblock {\em Physical Review Letters\/}, {\em 91\/}, 187904.

\bibitem[{Vaidman(1999)}]{vaidman1999}
Vaidman, L. (1999).
\newblock Variations on the theme of the {G}reenberger-{H}orne-{Z}eilinger
  proof.
\newblock {\em Foundations of Physics\/}, {\em 29\/}, 615--630.

\end{thebibliography}

\end{document}